\documentclass[12pt,preprint]{aastex}

\def\Re{\mathop{\rm Re}}

\shorttitle{TPF Coronagraph}
\shortauthors{Kuchner \& Traub}

\begin{document}

\title{A Coronagraph with a Band-Limited Mask for Finding Terrestrial Planets}
\author{Marc J. Kuchner\altaffilmark{1}}
\altaffiltext{1}{Michelson Postdoctoral Fellow}
\email{mkuchner@cfa.harvard.edu}
\author{Wesley A. Traub}
\email{wtraub@cfa.harvard.edu}
\affil{Harvard-Smithsonian Center for Astrophysics \\ Mail Stop 20, 60 Garden St., Cambridge, MA 02138}

\begin{abstract}
Several recent designs for planet-finding telescopes
use coronagraphs operating at visible wavelengths 
to suppress starlight along the telescope's optical axis while
transmitting any off-axis light from circumstellar
material.  We describe a class of graded coronagraphic image
masks that can, in principle, provide perfect elimination of
on-axis light, while simultaneously maximizing the Lyot stop
throughput and angular resolution.
These ``band-limited'' masks operate on the intensity
of light in the image plane, not the phase.  They can work with
almost any entrance pupil shape, provided that the entrance pupil
transmissivity is uniform, and can be combined with an apodized Lyot
stop to reduce the sensitivity of the coronagraph to
imperfections in the image mask.  We discuss some practical
limitations on the dynamic range of coronagraphs in the context of a
space-based terrestrial planet finder (TPF) telescope, and
emphasize that fundamentally, the optical problem
of imaging planets around nearby stars is a matter of
precision fabrication and control, not Fraunhofer diffraction
theory.
\end{abstract}

 
\section{INTRODUCTION}

The recent flood of indirect detections of extrasolar planets
\citep{exoplanets.org, schn01}\footnote{These websites about
extrasolar planets are available at  \url{http://exoplanets.org/} (Marcy et al. 2001)
and \url{http://www.obspm.fr/planets} (Schneider 2001).}
inspires us to search for an extrasolar planet that might be
capable of supporting human life.  Both NASA \citep{tpf}
and the European Space Agency \citep{darwin} have encouraged
research into the design of a space telescope specialized
for this purpose, a Terrestrial Planet Finder (TPF).

To detect an Earth-like planet orbiting a nearby star requires a
telescope capable of unprecedented dynamic range.
Initial TPF designs have concentrated on large-baseline
interferometers operating in the mid-infrared, where the contrast
between an earth-like planet and a sun-like star is minimal,
about $10^{-6}$ \citep{tpf}.
The expected planet/star contrast is a more daunting $10^{-10}$
at visible wavelengths, i.e., 0.3--1.1 $\mu$m.

However, a TPF operating at visible wavelengths offers several
advantages.  At shorter wavelengths, a smaller
telescope can obtain the required diffraction-limited resolution;
a 10 m baseline at 0.5 $\mu$m yields the same resolution as
a 200 m baseline at 10 $\mu$m.  Optical detectors require
much less thermal control than infrared detectors, reducing the
need for onboard cryogen and thermal shielding; a visible-light
telecope may operate at room temperature, while a telescope
operating in the thermal infrared must be cooled to $\sim40$K.  
Conventional optical telescopes form images, while an infrared
interferometer coupled to a single mode of the radiation field
focuses all the light in the interferometer beam onto a single
detector pixel; this distinction makes an optical TPF relatively
immune to background noise from exozodiacal dust.
Finally, the pair of biomarkers,
O${}_2$ and O${}_3$, available at visible wavelengths,
appears to be more useful than
the single biomarker, O${}_3$, available in the 
mid-infrared \citep{trau01, desm01}.

The visible-light TPF designs proposed to date
may all be loosely classified as coronagraphs.
A traditional coronagraph is a device that sends a wavefront
through two focii, an image plane and a pupil plane,
before it forms the final image,
so occulting masks strategically placed in the light path
can control how the net instrument transfer function varies over the
image plane.  Not all TPF designs use occulting masks,
allowing them to possibly dispense with some optical
surfaces that a complete coronagraph would include.  However,
the formal description of the propagation of light through a
coronagraph also describes these mask-less designs.

A conventional coronagraph typically uses a telescope which
has not been optimized for coronagraphy, with
an ordinary circular or obstructed circular aperture.
The telescope focuses light onto an occulting mask which
is transparent except near the optical axis, where it
becomes opaque, to block the incoming on-axis light.  The
pupil is then re-imaged onto a Lyot stop, which is
transparent near its center, but opaque near its edge to
block light diffracted by the occulting mask \citep{lyot39}.

Unfortunately, to achieve the necessary dynamic range a few
diffraction widths from the optical axis with a
circular aperture and a traditional graded on-axis occulting
mask requires a Lyot stop that blocks $\gtrsim 60\%$ of
the entrance pupil.  To get around this stumbling block,
some visible-light TPF designs invoke specially shaped pupils
to separate starlight from planet light,
or pupils with nonuniform transmissivity \citep{nise01,sper01,kasd01}.
We refer to pupils with nonuniform transmissivity as apodized pupils.
Other coronagraph designs \citep{guyo01} involve masks that
must directly manipulate the phase of the light they transmit.

We show here that a conventional coronagraph with a
graded pupil plane mask can, in principle, provide arbitratily
large dynamic range without need for phase control or a severe Lyot stop.
We describe a class of occulting masks and matching Lyot stops
that can work in conjunction with any un-apodized
entrance pupil to completely block the light from an on-axis source.
Following this exposition, we consider some of the practical
limitations on the dynamic range of coronagraphs in the context of TPF.

\section{CORONAGRAPHY}
\label{sec:coronagraphy}

\notetoeditor{I couldn't get the boldface sigmas in the
exponents to be the right size.}

The basic theory of coronagraphy has been discussed elsewhere
\citep{noll76,wang88,malb96,siva01}, but we review it here to establish our notation.
Let us assume that the telescope primary has a size scale $D$,
and that we are working at a wavelength $\lambda$.  These two physical
scales will almost always appear in combination, so we will frequently use
the abbreviation $D_{\lambda}=D/\lambda$.
The coordinates in the pupil plane
will be ${\mbox{\boldmath $\sigma$}}=(u,v)$, and 
the coordinates in the image plane will be ${\bf r}=(x,y)$.
In general, the hat accent will denote Fourier conjugation, i.e.
$\hat X({\bf r}) = \int d \Omega \ X({\mbox{\boldmath $\sigma$}}) e^{-2 \pi i {\bf r} \cdot {\mbox{\boldmath $\sigma$}}}$.
Quantities without a hat reside in the pupil plane; quantities
with a hat reside in the image plane.  We will work primarily
in the pupil plane.

Consider an electromagnetic wave
incident on an optical telescope.  Explicitly, we would write that
the field is a vector with magnitude
\begin{equation}
E({\mbox{\boldmath $\sigma$}},z,t)=F({\mbox{\boldmath $\sigma$}}) \Re\left\{ e^{i(\omega t - kz)}\right\},
\end{equation}
where $k=2 \pi/\lambda$, $\omega$ is the wave's angular frequency, $t$ is time,
and the wave propagates in the direction of the $+z$-axis.
Hereafter, we will drop all factors of $\Re\left\{e^{i(\omega t - kz)}\right\}$, 
and we will refer to the incoming field as $F({\mbox{\boldmath $\sigma$}})$.
The intensity of light associated with this field is proportional to
$|F({\mbox{\boldmath $\sigma$}})|^2$.

The wave first encounters the telescope's primary mirror, 
which has aperture function $A({\mbox{\boldmath $\sigma$}})$, so that it transmits
a field $A({\mbox{\boldmath $\sigma$}}) \cdot F({\mbox{\boldmath $\sigma$}})$.  The
telescope then forms an image from this transmitted field.   Assuming Fraunhofer
diffraction applies, we can write the image field
as $\hat A({\bf r}) \star \hat F({\bf r})$, where $\star$ denotes convolution.

In a coronagraph, the image is focused on an occulting mask
with amplitude transmission factor (ATF) denoted by $\hat M({\bf r})$, and
intensity transmission factor (ITF) equal to $\left|\hat M({\bf r})\right|^2$.  A mask
ATF may be negative or even complex, but if a mask is to operate on
the intensity of light and not its phase, its ATF
must be real, $0 \le \hat M({\bf r}) \le 1 $.   The mask opacity is
$1 - \left|\hat M({\bf r})\right|^2$.

After the occulting mask, the field is
$\hat M({\bf r}) \cdot \left(\hat A({\bf r}) \star \hat F({\bf r})\right)$.
Successive optics in the coronagraph transform this product
to a second pupil plane, where the field is
$M({\mbox{\boldmath $\sigma$}}) \star \Bigl( A({\mbox{\boldmath $\sigma$}}) \cdot F({\mbox{\boldmath $\sigma$}})\Bigr)$.
Now the field passes through a Lyot
stop, which can be described by an aperture function $L({\mbox{\boldmath $\sigma$}})$.
The final field has an amplitude 
$L({\mbox{\boldmath $\sigma$}}) \cdot \Bigl( M({\mbox{\boldmath $\sigma$}}) \star \left( A({\mbox{\boldmath $\sigma$}}) \cdot F({\mbox{\boldmath $\sigma$}})\right)\Bigr)$.
This field is re-imaged onto a detector, which detects the image's intensity.

Notice that without an occulting mask ($\hat M({\bf r})=1$), the
coronagraph has a transfer function
$L({\mbox{\boldmath $\sigma$}}) \cdot A({\mbox{\boldmath $\sigma$}})$.  In
this case, the Lyot stop and the aperture stop are redundant and they
could be interchanged without affecting the final image.  However,
if $\hat M(r)$ is not a constant then interchanging the Lyot stop and the
aperture stop will generally alter the final image.

\section{A ONE-DIMENSIONAL EXAMPLE}
\label{sec:gmask}

We will illustrate how a coronagraph treats on-axis light by
following the propagation of an incoming plane wave
through a coronagraph.
For simplicity, we will treat a one-dimensional coronagraph,
and restrict ourselves to functions of $u$ and $x$.
For on-axis light, $F({\mbox{\boldmath $\sigma$}})$ is a
constant; for our one-dimensional example, we chose
$F(u)=1$.  We represent the telescope
pupil, $A(u)$, with a tophat aperture, where the
tophat function is $\Pi(u) = 1$ for $ -1/2 < u < 1/2$,
and $\Pi(u)=0$ elsewhere.

We prefer to visualize the operation of a coronagraph entirely in
the pupil plane.  Note that the mask ATF, $\hat M(x)$, which
multiplies the electric field, is an image plane quantity.  We
prefer to plot the conjugate of the mask ATF, $M(u)$,
a pupil-plane quantity.  

\begin{figure}
\epsscale{0.85}
\plotone{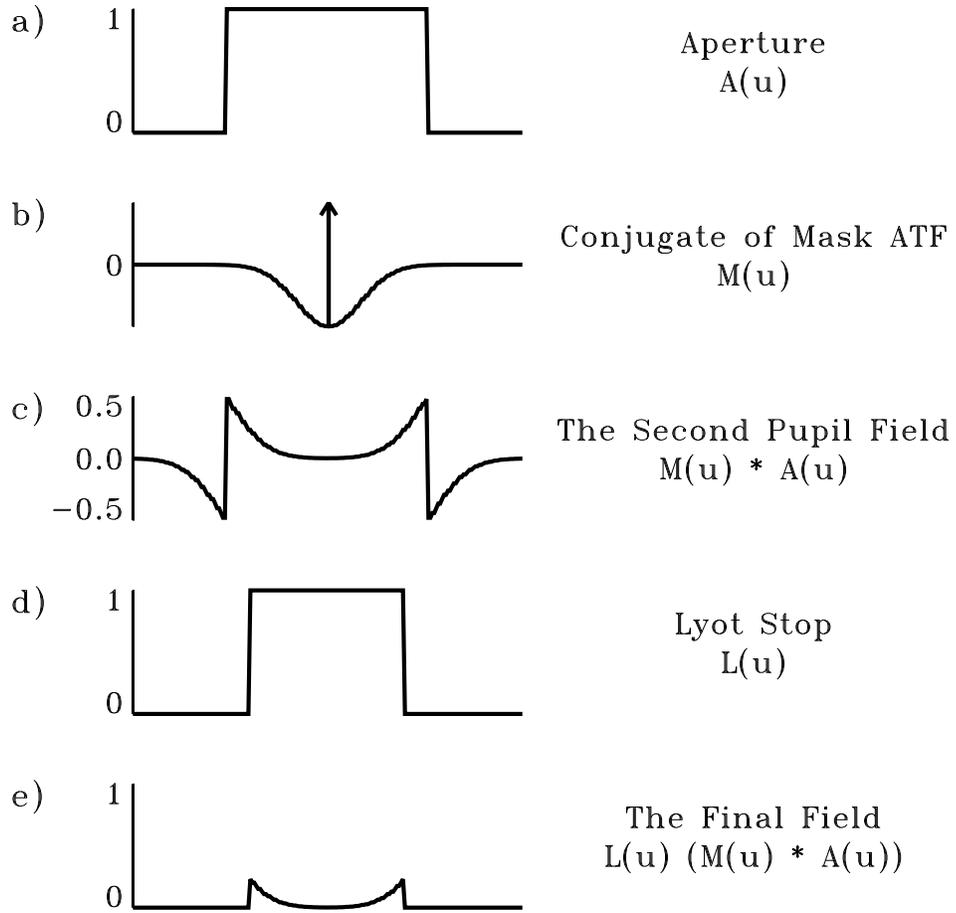}
\caption{One-dimensional coronagraph with a
Gaussian image mask, examined in the pupil plane.  First an
incoming field hits the primary aperture, then the image mask.
The mask ITF multiplies the image intensity;
in other words, the conjugate of the image mask ATF (b)
becomes convolved with the aperture function (a).
The result, (c), can be passed through a Lyot stop, (d),
leaving the final field (e).}
\label{fig:gaussianmask}
\end{figure}

Figure~\ref{fig:gaussianmask} illustrates the operation of 
a conventional one-dimensional coronagraph using the same example discussed
at length in \citet{siva01}.  Figure~\ref{fig:gaussianmask}a shows the incoming
field, multiplied by the telescope pupil.  Since the incoming field is $F(u)=1$,
this quantity is just $A(u) = \Pi(u / D_{\lambda})$.

In a conventional coronagraph, the occulting mask is a dark spot
at the optical axis.  We have chosen a
simple Gaussian mask ATF
\begin{equation}
\hat M(x) = 1- e^{-(1/2)(x D_{\lambda}/x_0)^{2}}
\end{equation}
to represent this dark spot.
The mask ITF associated with this function is
\begin{equation}
\left| \hat M(x) \right|^2 = \left|1- e^{-(1/2)(x D_{\lambda} /x_0)^2}\right|^2.
\end{equation}
Figure~\ref{fig:gaussianmask}b shows the conjugate of this
mask's ATF,
\begin{equation}
M(u) = \delta(u/D_{\lambda}) - {{x_0 \sqrt{2 \pi}} \over {D_{\lambda}}}e^{-(1/2)(2 \pi x_0 u/ D_{\lambda})^2},
\end{equation} 
where $\delta$ is the Dirac delta function.

For our on-axis source, 
the field in the second pupil plane is a sum of
error functions and a tophat
(Figure~\ref{fig:gaussianmask}c). 
\begin{equation}
M(u) \star A(u) = {\rm erf}\Bigl(2 \pi x_0 (u /D_{\lambda} - 1/2)\Bigr)
+ \Pi(u /D_{\lambda})
- {\rm erf}\Bigl(2 \pi x_0 (u /D_{\lambda} + 1/2)\Bigr),
\label{eq:sumoferfs}
\end{equation}
where 
\begin{equation}
{\rm erf}(x) = {1 \over \sqrt{2 \pi}}\int_{- \infty}^x e^{-{x'}^2/2} \,dx'.
\end{equation}
In the pupil plane, the effect of
the image mask is to diffract power to the edges of the 
pupil.  In the next stage of the coronagraph, the second
pupil field meets a Lyot stop, which is traditionally a hard-edged
aperture with diameter $D_{\rm Lyot} < D$, represented by the tophat function in
Figure~\ref{fig:gaussianmask}d, $L(u)=\Pi(u \lambda/ D_{\rm Lyot})$.
The Lyot stop blocks the power in the vicinity
of the edges of the primary aperture, though some
power leaks through the center.
Figure~\ref{fig:gaussianmask}e shows this final
field,  $L(u) \cdot \Bigl( M(u) \star A(u) \Bigr)$, which
propagates to the final image plane to form an image.  

The coronagraph has reduced the power in the final
residual field to a small fraction of the
incident on-axis power.  This fraction can be controlled by
decreasing the opening of the Lyot stop, $D_{\rm Lyot}$.
However, reducing the diameter of the Lyot stop
decreases the overall throughput and angular resolution of the
coronagraph.  To detect a terrestrial planet with a 10 m telescope
requires a dynamic range of $10^{10}$ for point sources a few
diffraction widths from the image of a bright star.  Our
numerical experiments show that attaining this dynamical
range in a two-dimensional coronagraph with a Gaussian image
mask requires masking roughly 60--70\% of the collecting area
of the telescope.

\section{BAND-LIMITED MASKS}
\label{sec:blmask}

What if an occulting mask could be designed to place
all---not just most---of the diffracted power from an on-axis source
in the second pupil plane within narrow
zones of width $\epsilon D_{\lambda}$ near the sharp edges of
the entrance pupil?
Then combining this mask with a Lyot stop that blocked all
the light in those zones would block precisely all the power from
an on-axis source.  Such an occulting mask would have have a
mask ATF whose Fourier conjugate,
$M({\mbox{\boldmath $\sigma$}})$, would be zero everywhere
except in the narrow range
$|{\mbox{\boldmath $\sigma$}}| < \epsilon D_{\lambda}/2$.  In other words,
the mask would be band-limited.

\begin{figure}
\epsscale{0.85}
\plotone{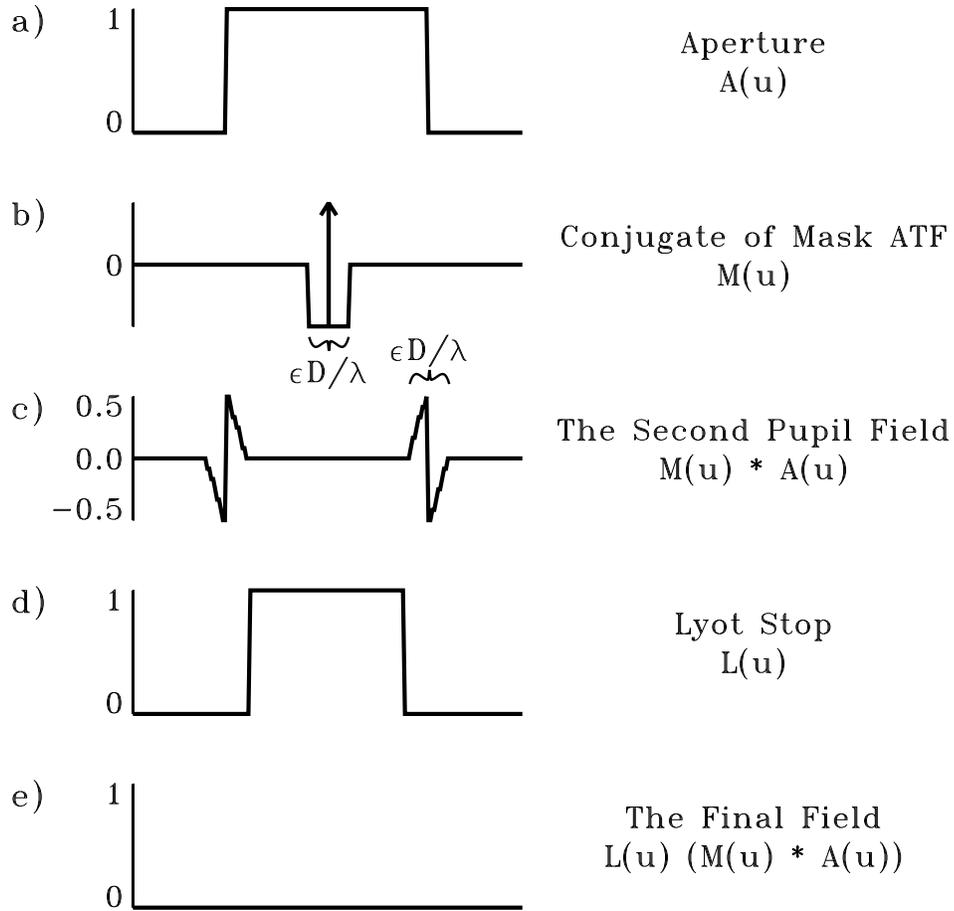}
\caption{Ideal band-limited coronagraph with
the same aperture (a) as the coronagraph in
Figure~\ref{fig:gaussianmask}.  Here the conjugate of
the mask ATF (b) is identically zero at all spatial
frequencies above some cutoff, $\epsilon D /(2 \lambda)$, so the second pupil
field (c) for an on-axis source is zero except within $\epsilon D/(2 \lambda)$
of the aperture edge.  Consequently, a Lyot stop, (d), can
block all the on-axis light, leaving zero final field (e).}
\label{fig:bwlmask}
\end{figure}

Figure~\ref{fig:bwlmask} illustrates the action of a one-dimensional
coronagraph with a band-limited mask.  
Figure~\ref{fig:bwlmask}a shows the same entrance pupil as in
Figure~\ref{fig:gaussianmask}a, a tophat aperture.  Again, we will
consider an on-axis incoming wave, $F(u)=1$.

We chose for this illustration the band-limited mask ATF
\begin{equation}
\hat M(x) = N(1 - \sin(\pi \epsilon D_{\lambda} x)/(\pi \epsilon D_{\lambda} x)),
\end{equation}
where N is a constant of normalization $N \approx 1/1.21723$ chosen
to keep $0 \le \hat M({\bf r}) \le 1$.
The ITF of this mask is
\begin{equation}
\left|\hat M(x)\right|^2 = N^2(1 - \sin(\pi \epsilon D_{\lambda} x)/(\pi \epsilon D_{\lambda} x))^2.
\end{equation}
The conjugate of this mask ATF is
\begin{equation}
M(u) = N\left(\delta(u/D_{\lambda}) - {1 \over {\epsilon D_{\lambda}}}\Pi(u / \epsilon D_{\lambda})\right), 
\end{equation}
as shown in Figure~\ref{fig:bwlmask}b.
This function is zero everywhere 
but where $|u| < \epsilon D_{\lambda}/2$.

The second pupil field, shown in Figure~\ref{fig:bwlmask}c, is now identically
zero except for within a small region around the edges
of the aperture, provided that the mask that is completely
opaque on-axis.  A simple Lyot stop (Figure~\ref{fig:bwlmask}d)
combined with a band-limited mask can eliminate all of the
light from an on-axis source.  The result is that the final field
from an on-axis source (Figure~\ref{fig:bwlmask}e) is identically zero.  

\begin{figure}
\epsscale{0.85}
\plotone{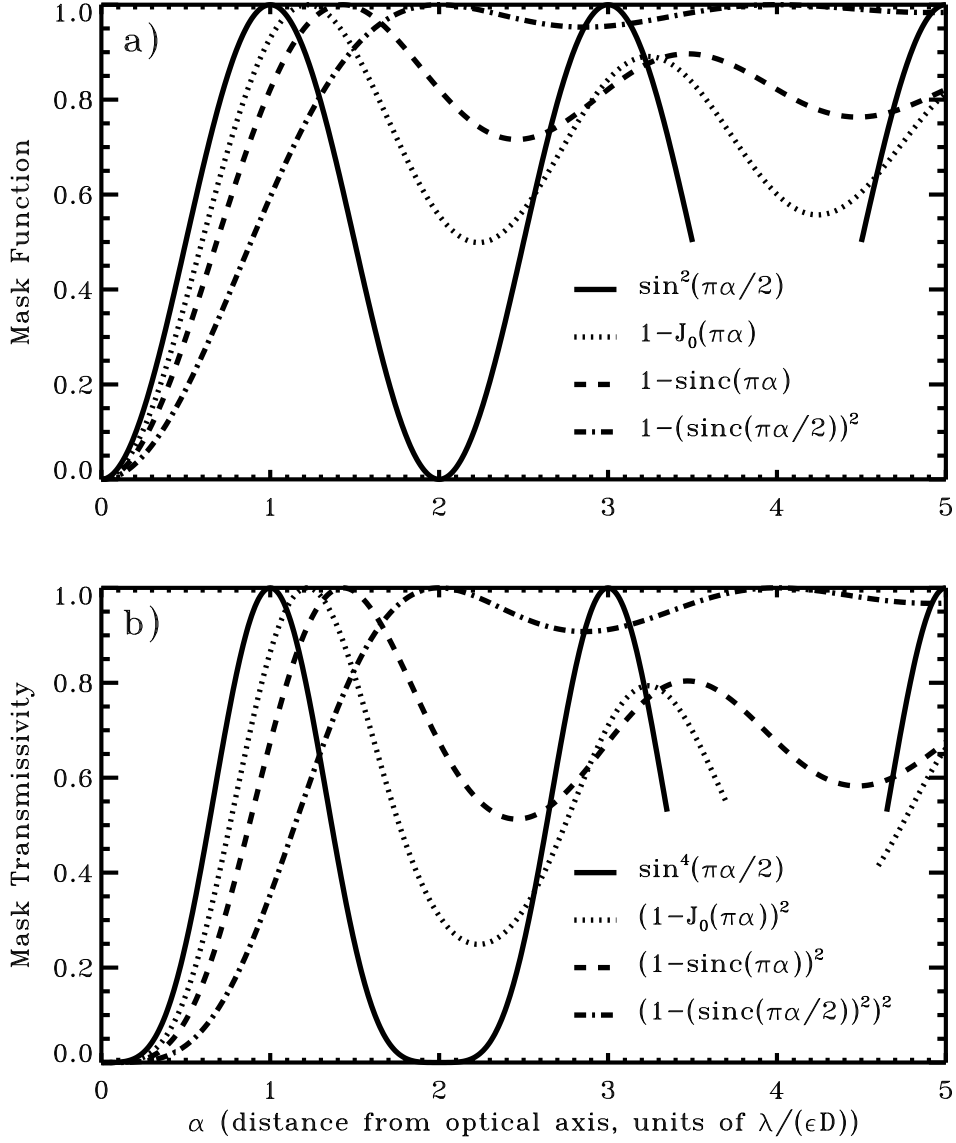}
\caption{Examples of band-limited functions that can
be used as mask ATFs.  a) A selection of mask ATFs
showing the fundamental tradeoff between the size of the
central core and the severity of the sidelobes.  b) The mask
transmissivities corresponding to these mask ATFs.}
\label{fig:examples}
\end{figure}

For a band-limited mask that operates on the intensity of
the first image, the mask ATF,
$\hat M({\bf r})$, may be any band-limited function of
${\bf r}=(x,y)$ with $0 \le \hat M({\bf r}) \le 1$, and $\hat M(0)=0$.
There are uncountably many such functions.
Figure~\ref{fig:examples} shows some examples of
one-dimensional band-limited functions that meet these criteria.
All of the functions shown have the
same bandwidth, $\epsilon$.  The ordinate in this figure
is distance from the optical axis in the image plane, measured
in units of $\lambda/(\epsilon D)$.  Figure~\ref{fig:examples}a
shows mask ATFs, and Figure~\ref{fig:examples}b shows the
corresponding transmissivities.  The range of functions in
this Figure illustrates a fundamental tradeoff; masks with small cores
have large sidelobes, and vise versa.

The functions shown in Figure~\ref{fig:examples} should be
viewed only as building blocks for constructing band-limited masks.
Any linear combination of band-limited functions
is a band-limited function.  Any product of band-limited
functions is a band-limited function, though since multiplying
functions corresponds to convolving their transforms,
the product of two mask ATFs will generally have
greater bandwidth than either
original function.  More options appear in two dimensions.
For example, the mask ATF of a
two-dimensional band-limited mask might
be a product of one band-limited function of $x$ with
another band-limited function of $y$, like
$\hat M(x,y)=(1-\cos{\pi \epsilon_u D_{\lambda} x})(1-\sin(\pi \epsilon_v D_{\lambda} y)/(\pi \epsilon_y D_{\lambda} y))$,
where $\epsilon_u$ and $\epsilon_v$ are the bandwidths in the
$x$ and $y$ (or $u$ and $v$) directions.
It is possible to design masks by combining simple
band-limited functions that are free of
sidelobes in specified regions of the image plane.

Naturally, constructing a band-limited mask requires a mask
of infinite extent. In reality, the physical size of the mask
will be limited to a few thousand diffraction widths. 
The result will be that $M({\mbox{\boldmath $\sigma$}})$
will be convolved by a sinc function, or possibly some other
taper, so a small amount of on-axis light will leak through
the coronagraph.

However, this limitation is not likely to be important.
For example, consider a one-dimensional coronagraph
whose mask ATF is tapered by a Gaussian 
$\exp((1/2)({\bf r} D_{\lambda}/ r_0)^2)$ with 
$r_0/D_{\lambda} \approx 100 \ \lambda/D$, so that the mask becomes opaque
at the edges. The tapering effectively convolves $M(u)$ with
$100\sqrt{2 \pi}\exp(-(1/2)(200 \pi u/D_{\lambda})^2)$.  The convolution of this function
with a tophat aperture falls off as
$1-\hbox{erf}(200 \pi (u/D_{\lambda}-1/2))$ near the
aperture edges (see Equation~\ref{eq:sumoferfs}).
For a Lyot stop designed to
block light within $\epsilon=0.2$ of the edges
of the entrance pupil, we find that the field at the
edge of the Lyot stop has been reduced by a factor of
$1-\hbox{erf}(40 \pi)$.  This corresponds to a 
reduction in transmitted power by a factor of better
than $(1-\hbox{erf}(40 \pi))^2$, a small quantity if
ever there were one.

\section{THE $\sin^4$ INTENSITY MASK, A NULLING INTERFEROMETER ANALOG}

\label{sec:nullingmask}

\begin{figure}
\epsscale{0.82}
\plotone{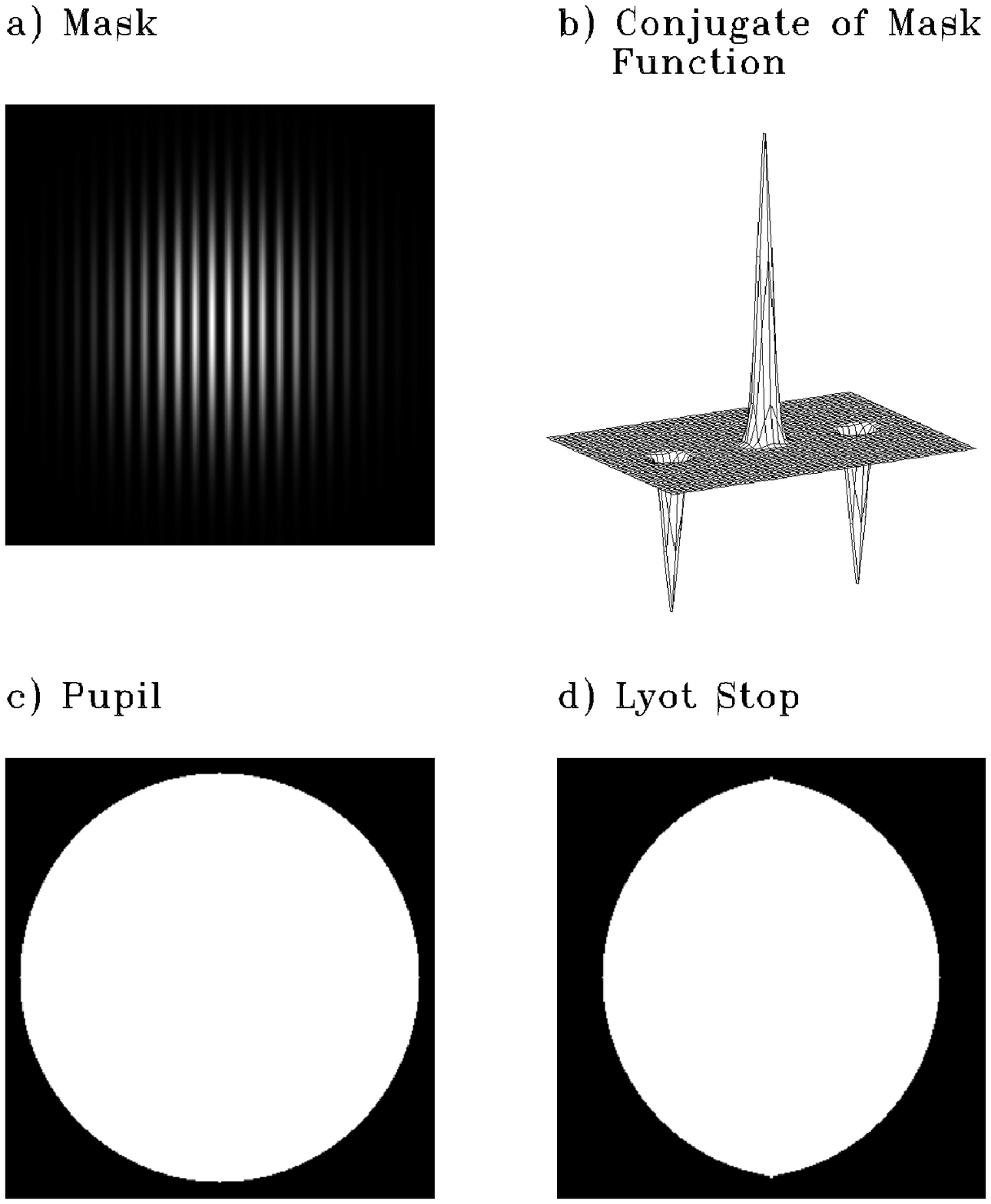}
\caption{The simplest band-limited mask is analogous
to a single-baseline nulling interferometer.  a) The mask
ITF, $\sin^4(x)$ multiplied by a slow taper.  Dark
areas are opaque.  b) The conjugate
of the mask ATF (Equation~\ref{eq:sinsquaredgaussm}).
This occulting mask can be used with any aperture shape, but
for the circular aperture shown in c), the corresponding
Lyot stop is d).}
\label{fig:nullingmask}
\end{figure}

A particularly simple example of a band-limited mask
has a mask ATF
\begin{equation}
\hat M({\bf r})={1 \over 2}-{1 \over 2}\cos(\pi \epsilon D_{\lambda} x)=\sin^2(\pi \epsilon D_{\lambda} x / 2)
\end{equation}
and ITF
$\left|\hat M({\bf r})\right|^2 = \sin^4(\pi \epsilon D_{\lambda} x/2)$.
The conjugate of this mask ATF is
\begin{equation}
M({\mbox{\boldmath $\sigma$}})=
-{1 \over 4} \delta({\mbox{\boldmath $\sigma$}}+\epsilon D_{\lambda} \tilde{\bf u})
+{1 \over 2} \delta({\mbox{\boldmath $\sigma$}})
-{1 \over 4} \delta({\mbox{\boldmath $\sigma$}}-\epsilon D_{\lambda}\tilde{\bf u})
\label{eq:sinsquaredm}
\end{equation}
where $\tilde{\bf u}$ is the unit vector along the $u$ axis.
Figure~\ref{fig:nullingmask} illustrates
a tapered version of this mask.  Figure~\ref{fig:nullingmask}a
shows the mask opacity, $1-|\hat M({\bf r})|^2$, where
\begin{equation}
\hat M({\bf r})=\sin^2(\pi \epsilon D_{\lambda} x/2) e^{-(1/2) {\bf r}^2/r_0^2},
\label{eq:sinsquaredgauss}
\end{equation}
and Figure~\ref{fig:nullingmask}b
shows the conjugate of the mask ATF,
\begin{equation}
M({\mbox{\boldmath $\sigma$}}) \propto
-{1 \over 4}e^{- (2 \pi r_0 / D_{\lambda})^2 ({\mbox{\boldmath $\sigma$}}+\epsilon D_{\lambda} \tilde{\bf u})^2/2}
+{1 \over 2}e^{- (2 \pi r_0 / D_{\lambda})^2 {\mbox{\boldmath $\sigma$}}^2/2}
-{1 \over 4}e^{- (2 \pi r_0 / D_{\lambda})^2 ({\mbox{\boldmath $\sigma$}}-\epsilon D_{\lambda} \tilde{\bf u})^2/2}
\label{eq:sinsquaredgaussm}
\end{equation}

The diffraction limit of a 10 m aperture at 0.5 $\mu$m is
10 mas.  A TPF design should be capable of 
finding planets as close as 30 mas from the star, an angular
separation corresponding to a transverse distance of 0.3 AU
at 10 pc.  With this design, we can search as
close as 30 mas from a star using a 10 m telescope at
$\lambda=0.5 \mu$m 
by choosing $\epsilon=0.21$, so the half-power point
in the mask ITF is at $3 \lambda/D$.  From the
prescription in Section~\ref{sec:blmask}
for a one-dimensional coronagraph, we know that
an appropriate Lyot stop would be one that blocks the region
within roughly $\epsilon D$ of the pupil edge.
This prescription applies in the $x$ and $y$ directions independently.
Since the bandwidth of this $\sin^4$ intensity mask is nearly
zero in the $y$ direction, (Figure~\ref{fig:nullingmask}b),
we can make the Lyot stop for this circular aperture
correspondingly wider in that direction, so the Lyot stop
only blocks $\sim \epsilon=21\%$ of the collecting area
(Figure~\ref{fig:nullingmask}d).

This mask has the smallest core of any band-limited
mask, and the price for this small core is that the
sidelobes are more than 50\% opaque over 64\% of the image
plane.  But this blockage need not be a setback.
Detecting a planet using a circular coronagraph with a
Gaussian mask will require rotating the telescope to distinguish
the image of a planet from artifacts produced by aberrations
in the optics; the artifacts will rotate with the telescope,
but the planet will not.  For a coronagraph
with a band-limited mask, the same rotation will provide
access to any discovery space blocked by the
mask.  Moreover, rotating the mask
independently of the rest of the optics to modulate the
transmitted light from the planet may provide a
further handle for distinguishing the image of a planet from
an optical aberration.

The $\sin^4$ intensity coronagraph is closely related to a
nulling interferometer. 
The transfer function of an interferometer
can be written $B({\mbox{\boldmath $\sigma$}}) \star A({\mbox{\boldmath $\sigma$}})$,
where $A({\mbox{\boldmath $\sigma$}})$ is the aperture
function for a single aperture, and 
$B({\mbox{\boldmath $\sigma$}})$ is a sum of delta functions that
represents the placement of the individual dishes.
If the dishes are centered at positions
${\mbox{\boldmath $\sigma$}}_i$ projected onto
the plane of the sky, then
$B({\mbox{\boldmath $\sigma$}})=\sum_{i} W_i \delta({\mbox{\boldmath $\sigma$}}-{\mbox{\boldmath $\sigma$}}_i)$,
where $W_i$ are weights applied to manipulate the
fringe pattern.  For example,
a conventional single-baseline interferometer may have $W_i=\{1,1\}$, so
$B({\mbox{\boldmath $\sigma$}})=\delta({\mbox{\boldmath $\sigma$}}-{\mbox{\boldmath $\sigma$}}_i) + \delta({\mbox{\boldmath $\sigma$}} + {\mbox{\boldmath $\sigma$}}_i)$.
But for a single-baseline nulling interferometer,
the beams are combined $\pi$ out of phase, so $W_i=\{1,-1\}$.
The $1:2:2:1$ nulling interferometer described
in appendix 1 of the TPF booklet \citep{tpf} has four dishes, weighted by
$W_i= \{-1/2,1,-1,1/2\}$.
The output signal from an imaging interferometer has intensity
\begin{equation}
I_{\hbox{interferometer}}({\bf r})\propto\left|\int e^{-2 \pi i {\mbox{\boldmath $\sigma$}} \cdot {\bf r}}
\Bigl(
B({\mbox{\boldmath $\sigma$}})
\star A({\mbox{\boldmath $\sigma$}}) \Bigr)
\cdot F({\mbox{\boldmath $\sigma$}}) \,du\,dv \right|^2.
\end{equation}
A Michelson interferometer generally couples only to the zero frequency
component of this signal, $I_{\hbox{interferometer}}({\bf 0})$.
By comparison, we saw in Section~\ref{sec:coronagraphy} that a coronagraph forms
an image 
\begin{equation}
I_{\hbox{coronagraph}}({\bf r})\propto\left|\int e^{-2 \pi i {\mbox{\boldmath $\sigma$}} \cdot {\bf r}}
L({\mbox{\boldmath $\sigma$}}) \cdot
\Bigl( M({\mbox{\boldmath $\sigma$}})
\star
\Bigl( A({\mbox{\boldmath $\sigma$}}) \cdot F({\mbox{\boldmath $\sigma$}}) \Bigr) \Bigr) \,du\,dv \right|^2
\end{equation}
If we set $F({\mbox{\boldmath $\sigma$}})=L({\mbox{\boldmath $\sigma$}})=1$,
the expressions for the detected intensities become identical in form, where $M({\mbox{\boldmath $\sigma$}})$,
the conjugate of the mask ATF, plays the same
role as $B({\mbox{\boldmath $\sigma$}})$, the
placement and weighting of the interferometer dishes.

\begin{figure}
\epsscale{0.85}
\plotone{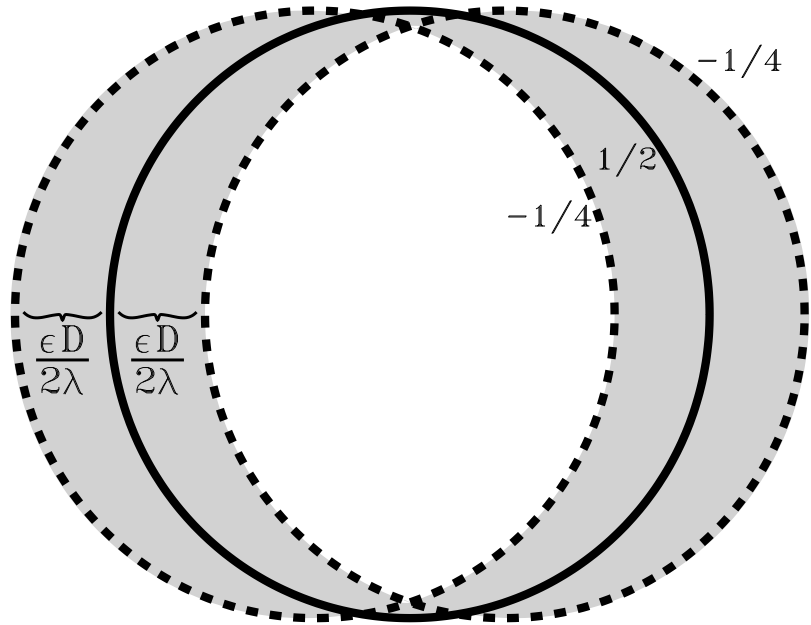}
\caption{Another way to visualize how the nulling
interferometer mask shown in Figure~\ref{fig:nullingmask} works.
Convolving the pupil transmissivity with the conjugate of
the mask ATF creates three virtual pupils, spaced
by $\epsilon D/(2 \lambda)$, weighted by -1/4, 1/2, and -1/4 respectively.
The nulling only works in the central region, colored white in this figure, where all three
pupils overlap, so the Lyot stop must block the region where
they do not overlap, colored grey.}
\label{fig:overlap}
\end{figure}

Figure~\ref{fig:overlap} illustrates the convolution of the conjugate
of the ATF, $M({\mbox{\boldmath $\sigma$}})$, for the $\sin^4$ intensity mask
(Equation~\ref{eq:sinsquaredm}) with 
the aperture function, $A({\mbox{\boldmath $\sigma$}})$, for a circular aperture.
The convolution effectively synthesizes three
apertures---interferometer dishes---with weights of $-1/4$, $1/2$, and $-1/4$,
spaced by $\epsilon D_{\lambda}/2$ wavelengths.  In Figure~\ref{fig:overlap},
circles indicate the three apertures, and the
grey areas indicate the regions illuminated by an on-axis source.

In an interferometer, the telescope apertures do not overlap.  But in
the $\sin^4$ coronagraph, the virtual apertures do overlap,
and only in the region where all three apertures
overlap does the field in the second pupil plane cancel to zero.
This region of overlap sets the shape of the Lyot mask
shown in Figure~\ref{fig:nullingmask}d.  Since any 
occulting mask ATF with even symmetry can be envisioned
as a weighted sum of $\sin^2$ mask ATFs, we conclude that
all intensity masks suffer from this imperfect overlap of virtual
pupils, forcing coronagraphs with intensity masks to
discard photons in the second pupil plane to achieve high
dynamic range.

Figure~\ref{fig:overlap} also demonstrates why a
band-limited mask does not work well
with an apodized entrance aperture.
A coronagraph effectively performs a weighted sum
of shifted copies of the entrance aperture with the weights chosen so 
one copy cancels other copies where they overlap.   If
the aperture does not have uniform transmissivity, it can not
generally uniformly cancel a shifted copy of itself (but note that
when the mask has zero bandwidth in one direction, the aperture
need not be uniform in that direction).

\section{ERRORS THAT CAN REDUCE A CORONAGRAPH'S DYNAMIC RANGE}

Generally, TPF designs aim to create a search area where
half of the photons from an on-axis source cancel the other half
of the photons from an on-axis source, leaving signals from off-axis
sources in the search area to fall on a detector relatively
un-attenuated.  An interferometer performs this
cancellation by interfering two beams.  A
coronagraph performs the same cancellation using diffraction.  
Consequently, a coronagraph may not circumvent the
requirements of accurate control over the phase
and amplitude of the wavefront that nulling interferometer
designs manifest.  Here we consider some of the noise sources
that can spoil a coronagraph's perfect cancellation of light from
an on-axis source.  These errors can reduce the dynamic range of
any coronagraph---with or without a band-limited mask.

\subsection{Off-Axis Light}

Before we proceed with our discussion of errors,
let us consider how a $\sin^4$ mask treats a point source, like
a terrestrial planet, at a small angle
${\bf \theta}=(\theta_x,\theta_y)$ from the optical axis. The
corresponding incoming electric field is
$F_s({\mbox{\boldmath $\sigma$}})=\exp(i \phi)$, where the
phase is $\phi = 2 \pi {\mbox{\boldmath $\sigma$}} \cdot {\bf \theta}/\lambda$.
In other words, the virtual pupils created by the mask are now associated with
phase gradients.

A phase gradient in the $y$ direction does not affect the
the weighted sum of the virtual pupils, as one might expect based on
the translational symmetry of the mask.
However, gradients in the $x$ direction will cause the right virtual
pupil in Figure~\ref{fig:overlap} to lag begin the central pupil by
$\Delta \phi = \pi \theta_x \epsilon D_{\lambda}$, and the left pupil to lead 
the central pupil 
by the same phase difference, $\Delta \phi$.  Totalling the three phasors
shows us that 
a uniform field of $\sin^2{\Delta \phi/2}F_p({\mbox{\boldmath $\sigma$}})$
will remain throughout the second pupil plane interior to the
Lyot stop.  The final image
plane will have an image of the point source, with power
suppressed by a factor of  $\sin^4{\pi \theta_x \epsilon D_{\lambda}/2}$
compared to the image an on-axis source would have in the
absence of the image mask.  In fact, the point source response
for an ideal coronagraph with any band-limited mask and properly chosen Lyot 
stop takes the form of the mask ITF times
the point-spread function created by the Lyot stop.

Recall that a Lyot stop reduces the signal
from a planet as well as the signal from an on-axis star.  It reduces the
effective collecting area, and spreads the planet's light over
a larger image plane area.  We shall say that the
coronagraph reduces the effective collecting area by a factor $f$, and
the peak brightness of the image of a point source by $f^2$.   For
a coronagraph with a Gaussian image mask with dynamic range
of $10^{-10}$ at $3 \lambda/D$, $f \approx 0.3$ in the the search area.  For
a comparable coronagraph with a $\sin^4$ band-limited mask whose
half power point is at $3 \lambda/D$,
$f \approx 0.8 \times \sin^2{\pi \theta_x \epsilon D_{\lambda}/2}$.

\subsection{Pointing Errors and The Finite Size of the Star}

A side effect of having off-axis light leak through the coronagraph is
that some starlight will leak through the mask due to a star's finite
size and any error in centering the star on the optical axis.
We can compute this leak by treating the stellar disk as a sum of
off-axis point sources.  Assume the star is a small 
disk with angular diameter $\theta_{\star}$ and uniform surface brightness
$B_{\star}=4 F_{\star}/(\pi \theta_{\star}^2)$, where
$F_{\star}$ is the total flux from the star.  Using the approximation
$\sin^4{\pi \theta_x \epsilon D_{\lambda}/2}  \approx (\pi \theta_x \epsilon D_{\lambda}/2)^4$,
we find that the fractional leak is
\begin{equation}
{1 \over {F_{\star}}}\int_{\rm disk} (\pi \theta_x \epsilon D_{\lambda}/2)^4 B_{\star} \, d\Omega
=(1/2048)(\pi \epsilon D_{\lambda})^4  \left[\theta_{\star}^4
+ 48 (\Delta \theta_x)^2 \theta_{\star}^2
+ 128 (\Delta \theta_x)^4 \right]
\end{equation}
where $\Delta \theta_x$ is the angular displacement of the center of the star
from the optical axis in the $x$ direction, the pointing error.  For example,
the Sun at a distance of 10 pc would have $\theta_{\star}=0.930$ milliarcseconds.
For a coronagraph with $D=10$ m, $\lambda=0.5 \mu$m, $\epsilon=0.2$ with no
pointing error, the fraction of the starlight that will leak through the mask to
form an image near the optical axis will be $5.0 \times 10^{-9}$.   With a pointing error
of 0.5 milliarcseconds along the $x$-axis, the fractional leak
increases to $1.3 \times 10^{-7}$.  The wings of the resulting extraneous image 
of the star will contaminate the search area; for instance a stellar leak due to
pointing error at the level of $10^{-7}$ may reduce the dynamic range of
the coronagraph to $10^{-10}$ in the search area.  Band-limited masks
with much less sensitivity to pointing errors and the finite size of the star can be
designed with the cost of decreased throughput.  One approach to pointing
a $\sin^4$ (or $\sin^6$, etc.) coronagraph that can also serve to distinguish
stellar artifacts from planets is to continually scan the mask across the
star to modulate the ratio between the transmitted starlight and the
light from any companion.

\subsection{Intensity Errors in the Primary Mirror}
\label{sub:intensity}

Suppose that a pupil of characteristic diameter
$D$ and real amplitude reflectivity factor $A_0$ with even symmetry
is covered with patches of diameter
$d$, so the ATF of the pupil is wrong by a real amount
$\Delta A$.  The corresponding ITF is
\begin{equation}
I = I + \Delta I = |A_0 + \Delta A|^2 = A_0^2 + 2 A_0 \Delta A + (\Delta A)^2
\label{eq:deltai}
\end{equation}
We can use this model to describe errors in the primary mirror reflectivity,
or errors in the apodizing function for an apodized pupil.  

The image of a point source formed by this pupil has intensity
\begin{equation}
\left|\hat A_0 + \widehat{\Delta A} \right|^2
= \left|\hat A_0 \right|^{2}
+ 2 \hat A_0 \Re\left\{\widehat{\Delta A}\right\}
+ \left|\widehat{\Delta A} \right|^2.
\end{equation}
The quantity $\widehat{\Delta A}$ is an image of the bad patches
on the pupil, a function which has characteristic radius $D/d$
diffraction widths, and area $\sim (D/d)^2$ squared diffraction widths.
This image is less intense than the image of the whole pupil
by a factor of $\sim (D/d)^2$, because the light is spread out into a
larger area than the image of the whole pupil.  The most potentially
dangerous errors are those which
concentrate their images in the search area, which is $\sim 4$
diffraction widths from the optical axis.
Let us focus on errors of this size scale, $d \approx D/4$.
For these errors, $|\widehat{\Delta A}|^2 \approx |\Delta A|^2 / 16$ or
$|\widehat{\Delta A}| \approx |\Delta A| / 4$.

If the pupil is a conventional circular aperture,
$|A_0|^2 \sim 10^{-3}$ in the search area.  In this case, the
term $2 A_0 \Re\left\{\Delta A\right\}$ dominates the error budget.
Since the Lyot stop reduces the area of the pupil that contributes to the halo
by a factor f, we require that $|2 \hat A_0 \Re\left\{\widehat{\Delta A}\right\}|f < 10^{-10} f^2 $,
or $|\Delta A| \lesssim 2 \times 10^{-7} f$.
If the pupil is a specially shaped or apodized aperture,  $|A_0|^2$ may be as low
as $\sim 10^{-10}$.  In this case, the quadratic term and the cross term are roughly
equal when the dynamic range is $\sim 10^{-10}$, and the requirement becomes
$|\Delta A| \lesssim 10^{-5} f$.

With these constraints on $|\Delta A|$, we can now ask, how big a $|\Delta I|$ can
we tolerate?  If the pupil is not apodized, the cross term in Equation~\ref{eq:deltai}
dominates everywhere on the pupil, and
$|\Delta I| \approx |2 A_0 \Delta A| \approx |2 \Delta A|$.  Then we require
$|\Delta I| \lesssim 4 \times 10^{-7} f $ for a circular aperture, and
$|\Delta I| \lesssim 2 \times 10^{-5} f$ for an aperture such as the one suggested by
\citet{sper01}.
If the pupil is apodized, then over the transmissive zones of the pupil, the
cross term dominates.  In the wings of the apodization function, where $|A_0| < 10^{-5}$,
the quadratic term in Equation~\ref{eq:deltai} may become
important.  These dark regions may be shaded less accurately than the rest of the surface.

\subsection{Wavefront Phase Errors}

If the primary mirror is not perfectly smooth it will deform the
reflected wavefront from an ideal shape.  Small perturbations will
create a halo of scattered light, which image masks and Lyot stops
can not correct, surrounding the image of a point source. 
This halo of speckles must be fainter than the core of the image
by a factor of $\sim 10^{-10}f^2$ so the halo
around a stellar image won't swamp the signal from a terrestrial
planet.   Suppose that the primary is actively corrected by a
two-dimensional array of pistons with $N_p$ pistons across the
diameter.  Let us assume for simplicity a circular aperture and
no mask or Lyot stop (so $f=1$) and ask how accurate the corrected
wavefront must be.

The distance between pistons in our array is $d_p = D/N_p$.
This array of pistons can nominally correct errors in
the wavefront reflected from the primary at spatial scales in the
pupil plane larger than $\sim 4d_p$ (assuming 2 pistons to
correct sine errors plus 2 pistons to correct cosine
errors).  Smaller scale errors will remain uncorrected.

Consequently, the active correction can reduce the intensity
of the scattered-light halo around a stellar image in the focal 
plane out to diameter $\theta_{halo} \simeq \lambda/(2d_p)$.
We must choose $N_p$ to be large enough so that the active
optics can eliminate the scattered light over the entire search
area.  Because we are concerned only with scattering diameters
less than $\theta_{halo}$, we will estimate the required maximum
r.m.s. wavefront error over only those scale sizes
between $4d_p$ and $D$, and ignore higher-frequency errors.

The Strehl ratio, $S$, of the image of a point
source is the ratio of the image's actual central intensity
to the central intensity the image would have if the
wavefront were perfect.  Since we are ignoring high frequency
errors which throw light into angles beyond $\theta_{halo}$,
we can say that the halo contains a fraction $ \sim 1-S$ of
the stellar photons.  Let us denote the nominal diameter of the
core image as $\theta_{core} = \lambda/D$.   Then the brightness
of the core of the image of a point source is
$\sim S / \theta_{core}^2$.  Likewise, if the halo is roughly
uniform, its brightness out to diameter $\theta_{halo}$ will be
$\sim (1-S)/\theta_{halo}^2$. 
 
The Strehl ratio, $S$, can be approximated by
$S = \exp(-\phi^2)$, where $\phi = 2\pi h/\lambda$, and
$h$ is the r.m.s. wavefront error.
The ratio of r.m.s. brightness in the halo to brightness
in the core is then
\begin{equation}
{{(1-S)/\theta_{halo}^2} \over  {S /\theta_{core}^2}} \simeq \left({{4\pi 
h} \over {\lambda N_p}}\right)^2 .
\end{equation}
To detect a terrestrial planet, this ratio
must be $< 10^{-10}$.  In other words, we require
$(4\pi h/\lambda N_p)^2 \lesssim 10^{-10}$.
For example, if $\lambda = 0.6\mu$m and $N_p=400$, the 
pistons must control the wavefront to an r.m.s. accuracy of $h 
\lesssim 2$ \AA.  

An active optics system often relies on
the imaged starlight to deduce the required corrections, so
the available starlight may limit the rate at which the corrections
can be calculated.  This condition makes achieving the tolerance mentioned
above even more daunting.  Achieving the required wavefront accuracy
is a major hurdle for coronagraph design \citep{malb95}. 


\subsection{Errors in the Image Mask Opacity}
\label{sub:stoptolerances}

 In the nulling interferometer, a
beam-combiner creates a nulling fringe, while in a coronagraph
with an occulting mask, the ``fringe'' is painted on an optical
surface.  This painted fringe must be fabricated precisely
just as the analogous interference fringe must be controlled
precisely.   This condition applies to both infinite-bandwidth
masks and finite-bandwidth masks.
For example, if an occulting mask is not completely
opaque in the center, but is has an ITF there of
$10^{-6}$, on-axis light will leak
through the Lyot stop at an intensity level of $10^{-6}$
on-axis, and an intensity level of a few
$\times 10^{-9}$ a few diffraction widths away, sufficient
to scuttle a terrestrial planet search. 
Because the light from the central star is concentrated
on a small part of the mask, only that small part of the
mask must be built to high precision, but the
required precision is high indeed.

Let us consider a more general perturbation to the
mask ATF, $\hat P({\bf r}) << 1 $.  The field from an on axis source in the
first image plane is $\hat A(u) \le 1$, so the field transmitted by the
mask is $\hat A (\hat M + \hat P)$, and the mask ITF is 
\begin{equation}
\left| \hat M + \hat P \right|^2
= \hat M^2 + 2 \hat M \hat P + \hat P^2 = \hat M^2 + \Delta ITF.
\label{eq:itrans}
\end{equation}
We will proceed by expressing $\hat P({\bf r})$ as a Fourier series.
Since the mask ATF for an intensity mask must be real, it is best to work in terms
of the sine and cosine basis functions, and write the perturbation
as
\begin{eqnarray}
\lefteqn{\hbox{\hskip -0.2 in} \hat P({\bf r})=P_0 + }& \nonumber\\
&\sum_{m,n} \biggl( P_{1m} \cos(\pi D_{\lambda} mx)
+ P_{2m} \sin(\pi D_{\lambda} mx)\biggr)
\biggl( P_{3n} \cos(\pi D_{\lambda} ny)
+ P_{4n} \sin(\pi D_{\lambda} ny)\biggr)
\label{eq:pexpand}
\end{eqnarray}

With Figure~\ref{fig:overlap} in mind, we recognize that the D.C. term,
$P_0$, creates a single virtual pupil centered on the optical axis, and
each sine and cosine term creates a pair of virtual pupils on
either side of the optical axis.  By analogy with the $\sin^4$ mask,
we can write the spacing between the two virtual pupils created by
a given term as $\epsilon_P D_{\lambda}$, where
$\epsilon_P = m$ or $n$ for the basis functions in Equation~\ref{eq:pexpand}.
In the case of the sine terms, the
fields from the virtual pupils cancel where the pupils overlap.   In the
case of the cosine terms, the fields from the virtual pupils do not cancel
themselves, but the D.C. term cancels them where they overlap, provided the
mask is precisely 100\% opaque in the center, as in the $\sin^4$
intensity mask.

\begin{figure}
\epsscale{0.79}
\plotone{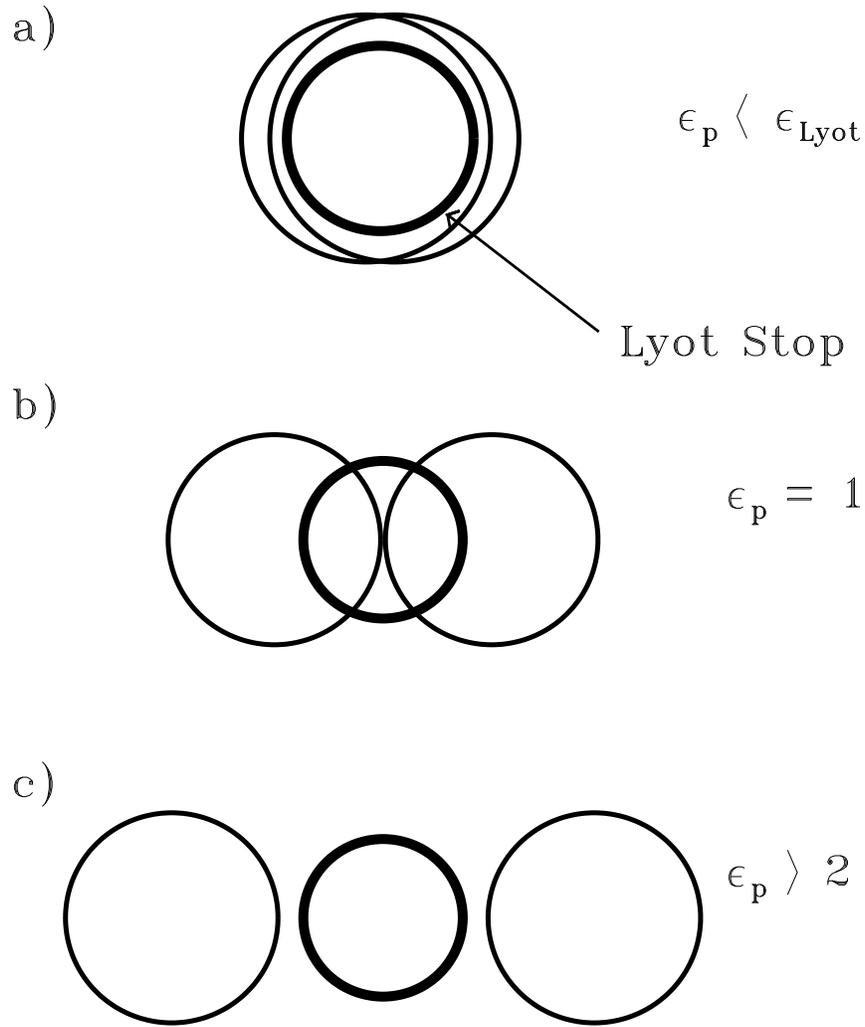}
\caption{Errors in the mask ATF create pairs of virtual
pupils, (the thin circles), which overlap with the Lyot stop
(the thick circle) if they have low spatial frequencies.
a) Low frequency errors ($\epsilon_p < \epsilon_{\rm Lyot}$) cancel to zero or can be balanced by the
D.C. term interior to the Lyot stop. b) Errors at
frequencies near the diffraction limit ($\epsilon_p \approx 1$)
do not cancel and can not be balanced by the
D.C. term within the Lyot stop. c) Errors with frequencies higher
than about twice the diffraction limit ($\epsilon_p > 2$) do
not propagate through the Lyot stop.}
\label{fig:noise}
\end{figure}

The perturbation wavenumbers fall into
three interesting regimes, shown in Figure~\ref{fig:noise}.  
For wavenumbers lower than the mask bandwidth, $\epsilon$, or rather,
the bandwidth blocked by the Lyot stop, $\epsilon_{\rm Lyot} \gtrsim \epsilon$,
in the appropriate direction, the virtual pupils overlap throughout
most of the region where the matched Lyot stop transmits.  In this
first regime (Figure~\ref{fig:noise}a), the Lyot stop suppresses
the effect of a sine perturbation.  The Lyot stop can suppress a
cosine perturbation too, provided that is comes with a matched D.C. term.
When $\epsilon_P \gtrsim \epsilon_{\rm Lyot}$, however, the Lyot stop
can not suppress the leak due to the perturbation.  The situation is
at its worst when the virtual pupils cease to overlap with
each other at $\epsilon_P = 1$.  This regime (Figure~\ref{fig:noise}b)
corresponds to perturbations on the size scale of the diffraction limit.  
When $\epsilon_P \gtrsim 2$, the virtual pupils are so widely separated
that they cease to overlap even with the Lyot aperture.
In this third regime (Figure~\ref{fig:noise}c), perturbations have no effect at all.

This analysis shows that the two potentially dangerous kinds of
errors are 1) the D.C. term, i.e., making the mask identically
opaque in the center, and 2) errors with size scales comparable
to the diffraction limit of the primary aperture.
For instance, there is no need to be concerned about errors in the mask ATF
caused by the finite number of molecules available for a mask coating.
However, even in the case of a mask which is
perfectly opaque in its center, an extraneous ripple in the
mask opacity on the scale of $\sim \lambda/D$ will add speckles to
the image of an on-axis point source in a halo whose radius is
$\sim 2 \lambda/D$.  Of course, these speckles
should be relatively easy to isolate; if the mask is rotated with
respect to the telescope, the speckles will rotate with the mask.

Now that we have decided that perturbations $\hat P({\bf r})$ on size scales
$\sim \lambda/D$ are the most dangerous, consider a mask with
intensity transmission factor $|\hat M({\bf r})|^2 + \Delta ITF({\bf r})$, where
$\Delta ITF({\bf r})$ is noise with size scale $\lambda/D$.
A single mode in $\hat P({\bf r})$ corresponds to perturbation in the
mask intensity transmission factor
$\Delta ITF({\bf r}) \approx 2 \hat M(r) \hat P(r)$, a function
which is a little bit more complicated than a plain
cosine or sine function but still has most of its power on the the same size scale,
if $\epsilon_P \sim \lambda/D$.  How big a $|\Delta ITF({\bf r})|$ can we tolerate? 

Consider a patch of a mask with size $\lambda/D$.  The flux that reaches
the patch from an on-axis source with
$F({\mbox{\boldmath $\sigma$}})=1$ is
$\sim (\lambda/D)^2 \left| \hat A({\bf r}_{patch}) \right|^2$.
A fraction $\Delta ITF({{\bf r}_{patch}})$ of this light is diffracted
into a halo with effective area $\sim (2 \lambda/D)^2$ in the final image.
This halo is the image of the region of overlap of some virtual pupils and
the Lyot stop; it may have wings or spokes that are just as bright as its
center.  The result is a noise background with an intensity
$\sim |\Delta ITF| \left| \hat A({\bf r}_{patch}) \right|^2 / 4$ in the search area.
We require that this noise background be less than
$10^{-10} f \left| \hat A({\bf 0}) \right|^2$.  In other words, if the
patch is near the center of the mask, where it is well-illuminated, the
patch must have $|\Delta ITF| \lesssim 4 \times 10^{-10} f$.  Elsewhere,
in the wings of the mask, where there is little illumination, the tolerance can be
much less severe.


\subsection{The Benefits of an Apodized Lyot Stop}

The best way to handle pointing errors, the finite angular
size of the star, and errors in the image mask opacity may
be to use an apodized Lyot stop.  Leak due to
pointing errors, the finite angular size of a star,
and a low-frequency errors in the image mask 
result in extraneous images of the star appearing near the
center of the final image plane.
Only the wings of these extraneous images interfere with
planet detection.
The leak due to these errors can be reduced by a factor
of $10^{-3}$ or more by combining a mask with a Lyot stop
designed to suppress the wings of the system's point spread
function (PSF).  For example, the Lyot stop could be gently apodized
with a Hanning window or some other graded function.   An apodized Lyot
stop can also limit the consequences of 
higher-frequency errors in the image mask, since these errors
produce virtual pupils that overlap the Lyot stop only at
the edges of the pupil plane, regions that are dark and graded in
an apodized Lyot stop.

Most of the on-axis light has been removed by the image mask
before it can reach the Lyot stop, so the construction tolerances
for this optic are much less severe than the tolerances computed
in Section~\ref{sub:intensity}.  Naturally, the Lyot stop still
must be completely opaque over the bandwidth of the light
diffracted by the image mask.  The cost for using such a specialized
Lyot stop will be reduced throughput.

\section{CONCLUSION}

We showed an example of a coronagraph design using a
mask with $\sin^4$ ITF that is analogous to
a nulling imaging interferometer with overlapped
entrance pupils.  Any coronagraphic occulting mask with even symmetry
can be decomposed into a weighted sum of these primitive masks.
If the sum is band-limited, with a reasonably
small bandwidth, a matched Lyot stop
can be designed to block identically all of the light from
an on-axis source while discarding a minimal fraction of
the light from an off-axis source.
The Lyot stop matched to the $\sin^4$ ITF mask
may use as much as $79\%$ of the collecting area of a
10-meter circular telescope to search for planets 30 mas
from nearby stars.  Planets imaged through such a
large-aperture Lyot stop benefit from a large fraction
of the collecting area and angular resolution of the
primary mirror.   Other band-limited masks may be
designed with much smaller sidelobes at the price of a
narrower Lyot stop.

Imaging extrasolar terrestrial planets
with a coronagraph requires very precise optics.  We pessimistically
restricted our error analysis to the case where all the power in the
errors is at the most dangerous size scales, where errors scatter light into a
halo with radius roughly a few $\lambda/D$.   In this case, we require
a beam from the primary mirror with r.m.s.~pathlength error of $\lesssim 1$ \AA\
and intensity errors of $\lesssim 10^{-10}$.  We also require 
an image mask with low-frequency
intensity transmission factor errors $\lesssim 10^{-10}$ within
a few diffraction widths of the optical axis.
An apodized pupil can be combined with in image mask to lighten the
error tolerances on both the image mask opacity and the apodizing
mask opacity if the apodized pupil is placed in the second pupil
plane as an apodized Lyot stop.  Building
a telescope that meets these tolerances may require substantial new developments
in optical technology.  However, before we can build such precision optics,
we can adapt the techniques described here for lower dynamic-range
applications on existing telescopes---perhaps to image
extrasolar giant plants.

\acknowledgements

We thank Anand Sivaramakrishnan and Eugene Serabyn for the name 
``Band-Limited Mask'' and we thank Rafael Millan-Gabet, John Monnier, and the
referee for close readings.

This work was performed in part under contract with the Jet
Propulsion Laboratory (JPL) through the Michelson Fellowship
program funded by NASA as an element of the Planet Finder
Program.  It was funded in part by Ball Aerospace
Corporation via contract with JPL for TPF design
studies.  JPL is managed for NASA by the California
Institute of Technology.

\end{document}